\documentclass[aps,prc,reprint, longbibliography,notitlepage,nofootinbib,nobibnotes,superscriptaddress,amsmath,amssymb,preprintnumbers]{revtex4-2}

\usepackage{subcaption}

\usepackage{graphicx} 
\usepackage{dcolumn}
\usepackage{bm}        
\usepackage{amssymb}   
\usepackage{adjustbox}  
\usepackage{amsmath,array}
\usepackage{comment}
\usepackage{tikz}
\usepackage{natbib}
\usepackage{MnSymbol}
\usepackage{caption}
\usepackage[force]{feynmp-auto}
\usepackage{subcaption}
\usepackage[colorlinks=true,linkcolor=blue,citecolor=blue, urlcolor=blue]{hyperref}
\usepackage[capitalise]{cleveref}
\usepackage     [utf8]                  {inputenc}
\usepackage     [T1]                    {fontenc}
\usepackage     [english]               {babel}
\usepackage{epigraph}
\usepackage{etoolbox}
\usepackage{ dsfont }
\usepackage{physics}
\usepackage{float}
\usepackage{xcolor}
\usepackage{siunitx}
\makeatletter
\patchcmd{\epigraph}{\@epitext{#1}}{\itshape\@epitext{#1}}{}{}  
\newcommand*\eqsize{%
\@setfontsize\mysize{9.0}{9.0}%
    }
    \usepackage{multirow}
\makeatother

\usepackage{xfrac}
\usepackage{amsmath,amssymb}
\usepackage{esdiff} 
\usepackage{commath} 
\usepackage{bbm} 
\usepackage{braket}
\usepackage{slashed}

\allowdisplaybreaks
\newcommand{\sNN}{s_{\mathrm{NN}}}

\newcommand{\rmd}{\mathrm{d}}

\newcommand{\Zr}{$^{96}_{40}$Zr}

\newcommand{\Ox}{$^{16}_{8}$O}
\newcommand{\Ar}{$^{40}_{18}$Ar}
\newcommand{\Cu}{$^{63}_{29}$Cu}
\newcommand{\Xe}{$^{129}_{54}$Xe}
\newcommand{\Au}{$^{197}_{79}$Au}
\newcommand{\Pb}{$^{208}_{82}$Pb}

\newcommand{\xT}{\mathbf{x}}

\newcommand{\sT}{\mathbf{s}}
\newcommand{\pT}{\mathbf{p}}
\newcommand{\qT}{\mathbf{q}}
\newcommand{\kT}{\mathbf{k}}

\newcommand{\SNN}{\sqrt{s_{\mathrm{NN}}}}

\newcommand{\Dipper}{\textsc{McDipper}}

\definecolor{oscar}{RGB}{22, 156, 172}

\begin{document}

\date{\today}

\title{ Baryon stopping and charge deposition in heavy-ion collisions due to gluon saturation}

\author{Oscar Garcia-Montero}
\email{garcia@physik.uni-bielefeld.de}
\affiliation{Fakult\"at f\"ur Physik, Universit\"at Bielefeld, D-33615 Bielefeld, Germany}

\author{Sören Schlichting}
\affiliation{Fakult\"at f\"ur Physik, Universit\"at Bielefeld, D-33615 Bielefeld, Germany}

\begin{abstract} 
 We compute baryon and electric charge deposition in high-energy heavy-ion collisions using the Color Glass Condensate (CGC) Effective Field Theory, where at leading order charge is deposited through multiple scatterings of valence quarks with a saturated gluon target. A simplified phenomenological formula is derived to describe charge deposition, from which the parametrical dependence with collisional energy and geometry can be extracted. We present an approximate analytical prediction of the so-called baryon stopping parameter $\alpha_B$, which shows excellent agreement with the state-of-the art extractions of $\alpha_B$ from experimental data. These results are further validated using the {\Dipper} framework, by computing charge deposition at midrapidity across a range of collision energies ($\SNN = 62.4 - 5020$ GeV).  
 
\end{abstract}

\maketitle

\section{Introduction}

In heavy-ion collisions a large number of protons and neutrons collide into each other, forming a deconfined state of quarks and gluons, the so-called Quark Gluon Plasma (QGP). Net baryon number, $B$, as well as net electric charge, $Q$, are conserved  quantities in nature which means that the total integrated charges are also conserved during heavy-ion collisions (HIC). During the very first instants of the collision, these conserved quantities are distributed throughout the newly formed QGP, leaving imprints across the transverse plane, with non-trivial longitudinal structures~\cite{ATLAS:2017rij,Bozek:2015bna,Petersen:2011fp,Nie:2019bgd,CMS:2015xmx}. Even though the densities of all conserved quantities evolve via advection and diffusion throughout the lifetime of QGP; the longitudinal structures observed in the final state are believed to be intimately connected to the initial conditions~\cite{Savchuk:2023yeh,Fotakis:2019nbq}. Hence, to understand the production of particles with these two quantum numbers, it is essential to understand the mechanism for charge deposition in the initial stage of the collision. 

Developing a first principles understanding of the initial net charge deposition is particularly relevant to the Beam Energy Scan (BES) program at RHIC~\cite{Tlusty:2018rif}, which focuses on the creation of hot nuclear matter at non-vanishing baryon charge, which is achieved by colliding heavy ions at low and intermediate energies. These collisions explore different trajectories in the QCD phase diagram, effectively sweeping over it to find signatures of the QCD critical point and/or a first order QCD phase boundary. While these searches are performed at center-of-mass erngies $\sqrt{s}_{NN} \leq 200 {\rm GeV}$, they can be complemented through searches at higher energies but quite forward/backwards rapidities, where baryon densities are also expected to be high~\cite{Brewer:2018abr,Li:2018ini,Du:2023gnv,Li:2023kja}. One advantage of such high-energy forward searches is that models describing the QCD matter deposition into the fireball are in general better constrained theoretically. The advent of next generation LHC detectors such as the Forward Calorimeter in ALICE (FoCal), the major upgrade ALICE 3 and the upgrades in settings of ATLAS, CMS and LHCb will allow the community to explore the physics of the large rapidity  regions further,~\cite{Arslandok:2023utm,ALICE:2020mso,ALICE:2022wwr}. Therefore, understanding baryon stopping at intermediate and high collisional energies becomes a necessity. 

Net baryon or -- as a proxy -- net proton rapidity distributions reveal information about the baryon stopping mechanism, i.e. how nucleons -and the associated baryon number-  is decelerated from the initial beam rapidity towards the mid-rapidity region. By looking at these distributions along different energies \cite{NA49:1998gaz,BRAHMS:2003wwg}, one observes the appearance of a double-peaked structure with long tails towards the midrapidity region, with the maxima of baryon number deposition shifting to higher rapidities with increasing energy (c.f. \cref{fig:diagram}). It is precisely this shift of the distributions, and therefore of the midrapidity tails which effectively causes the decrease in deposition power at midrapidity with increasing center of mass energy $\sNN$.
In the literature, a diverse array of mechanisms of baryon stopping have been discussed, including hadronic transport \cite{Mohs:2019iee,Schafer:2021csj,Weber:2002qb}, Markovian processes in QCD~\cite{Hoelck:2020iei} as well as mechanical deceleration mechanisms of string dynamics~\cite{Shen:2017bsr}. In recently years, the so called baryon junction, a topological configuration of gluons inside baryons \cite{Kharzeev:1996sq} has received increasing attention, due to its potential in accounting for certain features of baryon charge deposition in high-energy collisions~\cite{Shen:2022oyg,Frenklakh:2024mgu}. Also in recent  literature, this mechanism has been contrasted exclusively to the deposition of charge through perturbative scattering of valence quark from the projectile with the target. It is important to note that this a limit used traditionally by event generators such as Pythia~\cite{Andersson:1983ia,Sjostrand:2006za}, POWHEG~\cite{Frixione:2007nw} and HERWIG\cite{Corcella:2000bw}, which fails to account for the observed baryon charge deposition, as brought to attention in a recent paper by Lewis \textit{et al.} (see Ref.~\cite{Lewis:2022arg}). 

In this paper, we present an alternative picture of net-charge deposition in heavy-ion collision, by re-imagining the perturbative quark scattering in the context of a saturated target. For this, we computed charge deposition using the Color Glass Condensate Effective Field Theory (CGC EFT) at leading order (LO), where $B$ and $Q$ are deposited via multiple scatterings of single valence quark with a densely populated gluon target~\cite{Dumitru:2002qt}. 
We present a simplified formula, derived from the single quark production which can easily be adapted to other phenomenological approaches, such as e.g. the TrENto initial state model~\cite{Ke:2016jrd}, or MC-Glauber 3D~\cite{Zhao:2022ugy} and thus be used for phenomenological purposes. Based on a reasonable set of assumptions, the net quark density can be computed as the sum of two contributions, 
\begin{equation}
\left(n_{q_f}\tau\right)_0=\left(n_{q_f}\tau\right)_{0,A\to B}+\left(n_{q_f}\tau\right)_{0,B\to A}
\label{eq:param_form_1}
\end{equation}
corresponding to the -forward- charge deposited from the quarks $q_{f}$ of nucleus $A/B$  scattering off the saturated target $B/A$. Each individual contribution is given by
\begin{equation}
	\begin{split}
		\left(n_{q_f}\tau\right)_{0,A\to B}=&  \sum_{h=p,n} x_{1}q_{v_{f}/h}\big(x_{1},Q_{s,B}^2(x_2,T_{B}(x))\big)\; T_{h/A}(\xT)  
  \label{eq:param_form_2}
	\end{split}
\end{equation}
where $x_{1}q_{v_f/h}$ is the collinear parton distribution of the valence quarks of flavor $f$ in hadron $h$, and $T_{h/A}$ denotes the thickness function of the hadron $h$  in nucleus $A$. Due to kinematics the longitudinal momentum fractions $x_{1/2}$ are approximately given by
\begin{equation}
x_{1/2}\simeq Q_{s}^{B}/\sqrt{s_{NN}} e^{\pm\eta}
\end{equation}
with the saturation scale $Q_{s,B}$ of nucleus $B$ self-consistently determined from a phenomenological parametrization $Q_{s,B}^{2}(x_2,T_{B}(x))$. The beauty of this formula is that the kinematic variables and the scale at which we are probing the quarks are fully determined by the characteristic scale of the gluon distribution of nucleus $B$, namely the saturation scale $Q_B^2$.
Based on this formula we obtain an analytic expression for the behavior of charge deposition at midrapidity, which is consistent with the trends in experimental data~\cite{Lewis:2022arg}. We further investigate the behavior of baryon deposition using numerical event-by-event studies of the {\Dipper}~\cite{Garcia-Montero:2023gex,Garcia-Montero:2023opu} intitial state, which is based on the $k_T$-factorized limit of the CGC. We explore the systems size and collisional energy dependence, finding promising results for the latter. 

This paper is organized as follows: in \cref{sec:dipper} we introduce the 3D quark charge stopping mechanism as given in the {\Dipper} model. Additionally, we derive, based on kinematic arguments a simplified formula which can be used parametrically for initial charge deposition in phenomenological studies. In \cref{sec:Results} we present results for initial baryon and electric deposition using the {\Dipper} code. More specifically, we discuss baryon stopping as a function of center-of-mass energy in \cref{sec:baryon}, while exploring its dependence on system size in \cref{sec:systemsize}. We finalize with our summary and conclusions.

\section {3D Baryon and Electric charge deposition in the {\Dipper} }
\label{sec:dipper}
Below we discuss the charge deposition in the so-called hydrid formalism of the CGC EFT~\cite{McLerran:1993ni,McLerran:1993ka,McLerran:1994vd} as implemented for valence quarks in the {\Dipper} initial state model~\cite{Garcia-Montero:2023gex}. Within this framework net charge density is deposited via multiple scatterings of valence quarks in a dense gluon field. The average net quark charge density $n_{q_f}$  can be extracted from the single (anti-)quark distributions, 
\begin{equation}
	\begin{split}
	(n_{q_f}\tau)_0 &=\int d^2\pT~\left[\frac{dN_{q_f}}{d^2\xT d^2\pT dy} - \frac{dN_{\bar{q}_f}}{d^2\xT d^2\pT dy}\right]_{y=\eta_s} \\
	&=\int d^2\pT~\left[\frac{dN_{q_{f,V}}}{d^2\xT d^2\pT dy} \right]_{y=\eta_s} \,.
	\end{split}
		\label{eq:QuarkCharges}
\end{equation}
where at LO in the CGC power counting, the valence quark yield is given by~\cite{Dumitru:2002qt,Dumitru:2005gt}
\begin{equation}
	\begin{split}
		\frac{dN_{q_{f,V}}}{d^2\xT d^2\pT dy} &= \frac{x_{1}q_{v_f/A}(x_{1},\pT^2,\xT)~D_{\rm fun}^{(B)}(x_2,\xT,\pT)}{(2\pi)^2} \\
		&+ \frac{x_{2}q_{v_f/B}(x_{2},\pT^2,\xT)~D_{\rm fun}^{(A)}(x_1,\xT,\pT)}{(2\pi)^2}\,.
	\end{split}
	\label{eq:single_quark_density}
\end{equation}
with the kinematic variables $\{x_i\}$ are given by 
\begin{equation}
x_{1/2}=\frac{\pT}{\sqrt{s_{NN}}} e^{\pm y}
\end{equation}
The result in \cref{eq:QuarkCharges,eq:single_quark_density} is a result of  the linear dependence of the charge deposition on the nuclear (anti-) quark parton distribution in the single quark production formulas at LO, as expressed in  as~\cite{Dumitru:2002qt,Dumitru:2005gt}\footnote{Note that this is true for the average charge deposition, i.e. as long as event-by-event fluctuations of the quark content of colliding hadrons are not taken on account.}. Since in the hybrid approach, baryon stopping is modeled as quark momentum broadening (and hence rapidity loss) through multiple scatterings with a low-$x$ target, there is an implicit requirement that the CGC degrees of freedom in target are, in fact low-$x$. One may be concerned, that, as an example, in the case of forward charge deposition, more backward rapidities lead to a larger effective value of $x_{\rm target}$, making this assumption questionable. However, we have checked explicitly that in such cases, the relative contribution to the deposition of energy of conserved charges is almost negligible.

\begin{figure}[t]
	\includegraphics[scale=0.24]{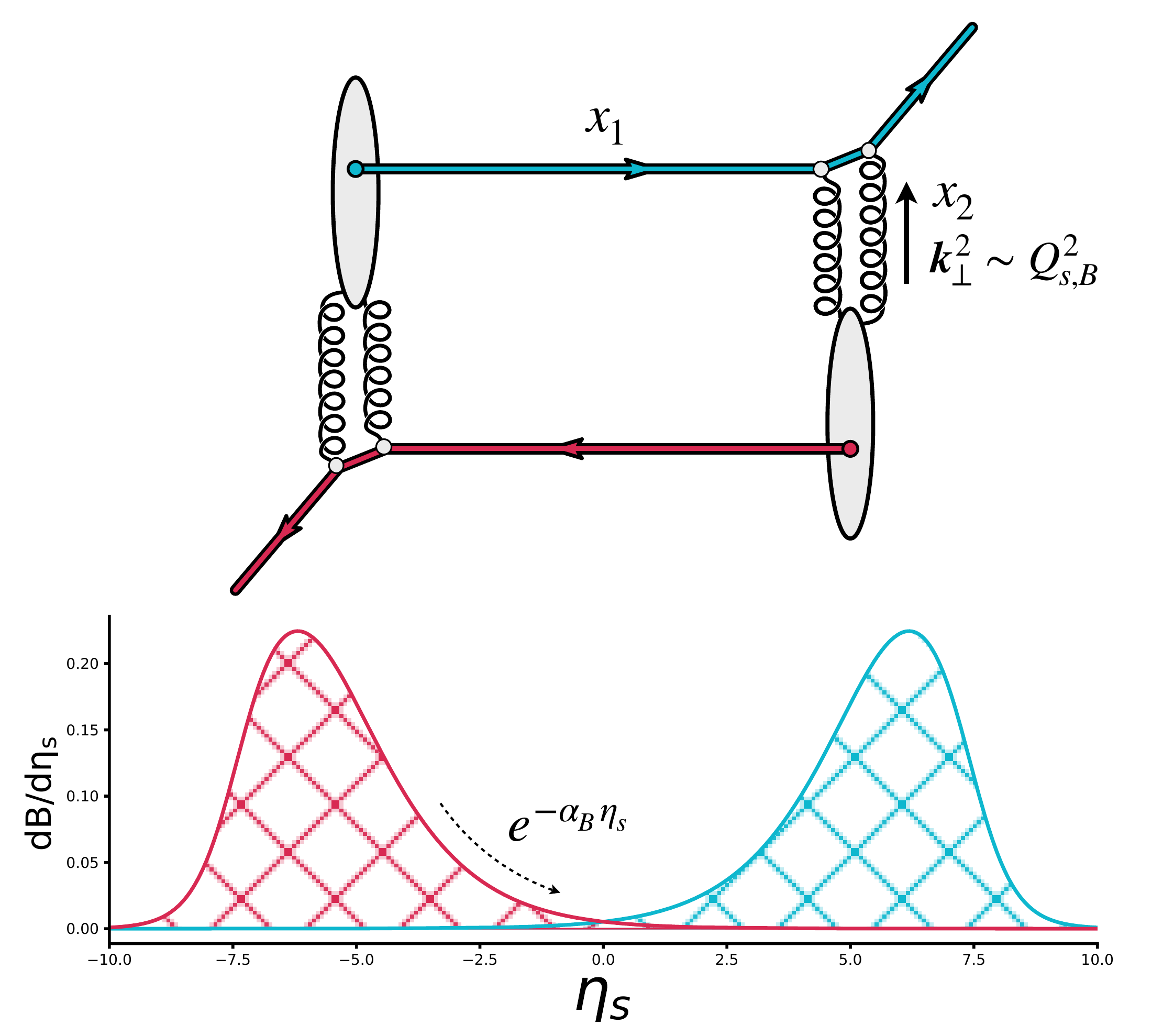}
	\caption{Charge deposition mechanism through baryon stopping in the saturation regime.}
	\label{fig:diagram}
\end{figure}

Intuitively,  \cref{eq:single_quark_density} states that valence charge is deposited into the interacting medium via the deflection of a collinear quark through multiple scatterings ~\cite{Mantysaari:2015uca,Lappi:2012nh} as illustrated schematically in \cref{fig:diagram}. Different terms in \cref{eq:single_quark_density} correspond to the stopping of valence charge contained in nucleus $A$ or $B$; the first one refers to the valence quarks from the right moving nucleus ($A$) being deflected towards the forward cone, while the second term refers to left moving collinear quarks from nucleus $B$, being deflected to the backwards cone. While this may look symmetric (in the case of same-species collisions ), as soon as there are fluctuations, charge deposition  will fluctuate on an event-by-event basis depending on the positions of the nucleons.

In this work, the collinear quark nuclear PDFs, $q_{v_f/A}(x_{1/2},Q^2,\xT)$, are given by scaling the independent nucleonic PDFs, with the nuclear thickness functions $T_{p/n}(\xT)$ of protons/neutrons, i.e.
\begin{equation}
	\begin{split}
		u_A(x,Q^2,\xT)&=u_{p}(x,Q^2)\; T_{p}(\xT) + u_{n}(x,Q^2)\; T_{n}(\xT)\,, \\
		d_A(x,Q^2,\xT)&= d_{p}(x,Q^2)\; T_{p}(\xT) + d_{n}(x,Q^2)\; T_{n}(\xT)\,, \\
		s_A(x,Q^2,\xT)&= T_{p}(\xT)\,s_{p}(x,Q^2) + T_{n}(\xT) \,s_{n}(x,Q^2)\,. 
		\label{eq:PDFS-pn}
	\end{split}
\end{equation}

As in Ref.~\cite{Garcia-Montero:2023gex}, we will assume isospin symmetry, as the parton distributions of the neutron are not well constrained.  This in turn  means that the neutron PDFs are set to the proton PDFs after the transformation $u\leftrightarrow d$, i.e. $u_{p/n}=d_{n/p}$, such that Eq.~\eqref{eq:PDFS-pn} simplifies to
\begin{equation}
	\begin{split}		u_A(x,Q^2,\xT)&=u(x,Q^2)\; T_{p}(\xT) + d(x,Q^2)\; T_{n}(\xT)\,, \\
		d_A(x,Q^2,\xT)&= d(x,Q^2)\; T_{p}(\xT) + u(x,Q^2)\; T_{n}(\xT)\,, \\
		s_A(x,Q^2,\xT)&= \left( T_{p}(\xT) + T_{n}(\xT) \right)\,s(x,Q^2)\,. 
		\label{eq:PDFS-pn-iso}
	\end{split}
\end{equation}

In this work we don't include nuclear effects into the quark distributions, as we expect the valence quark distributions to not receive large corrections~\cite{Eskola:2021nhw}. 
However, it is important to note that these effects can readily be included in the {\Dipper} through the inclusion of nuclear PDFs, readily available through LHAPDF~\cite{Buckley:2014ana}.

Finally, the last ingredient of the charge deposition is the fundamental dipole gluon distribution, $D_{\rm fun}^{(A/B)}$ of nucleus $A/B$,
	\begin{equation}
		D_{\rm fun}(x,\xT,\qT)=\frac{1}{N_c} \int_{\sT} \tr_{\rm fun}[U_{\xT+\sT/2} U^{\dagger}_{\xT-\sT/2}]~e^{\rm i \qT\cdot\sT} 
		\label{eq:dipole}
	\end{equation}
 which corresponds to scattering cross section of a quark in the Color Glass Condensate formalism of high-energy scattering.

\subsection{Parametrical formula}
\label{sec:paramform}
Now that we have established the theoretical basis, the main objective of this subsection is to develop a simple and intuitive formula that can be used for phenomenological purposes. For this, we will use the simplest model available, the Golec-Biernat Wusthoff (GBW)  model~\cite{Golec-Biernat:1999qor,Kowalski:2003hm,Kowalski:2006hc}, where the fundamental dipole gluon distribution is given by  a Gaussian distribution around the saturation scale,
	\begin{figure*}[t]
	\includegraphics[scale=0.55]{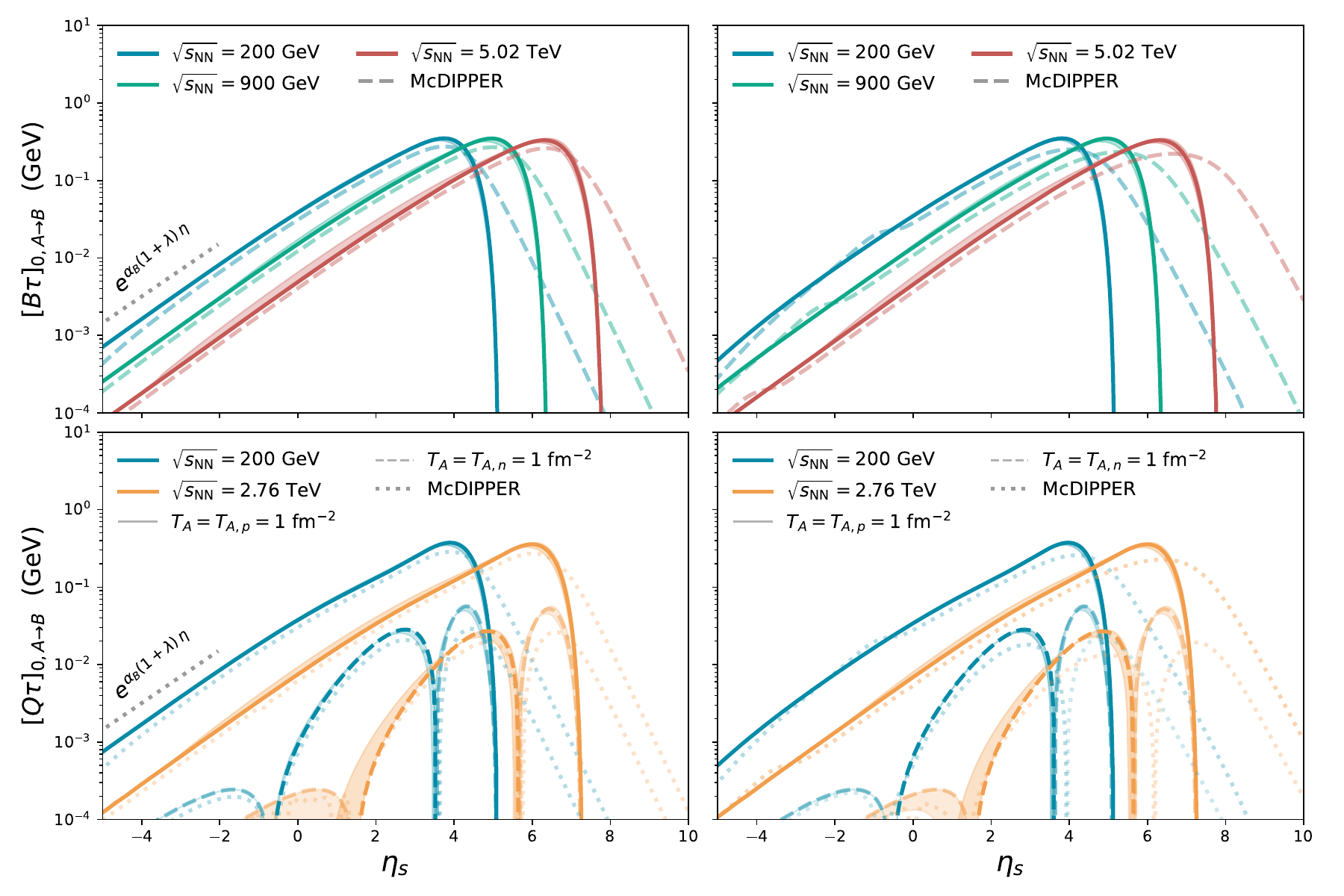}
	\caption{Comparison of baryon (\textit{upper panels}) and electric charge deposition (\textit{lower panels}) for two different energies, $200$~GeV and $2.76$~TeV. The (\textit{left}) panels correspond to the gaussian GBW model, while the (\textit{right}) panels correspond to the IP-Sat model, where the effective $Q_s^2$ is extracted from the dipole function as explained in the main text. }
	\label{fig:analytical_charge}
\end{figure*}

\begin{equation}
	D_{\rm fun}(\xT,\qT)=\frac{4\pi}{Q_F^2(x,\xT)} \exp\left[-\frac{\kT^2}{Q_F^2(x,\xT)}\right]\,.
	\label{eq:GBWDip}
\end{equation}
where $Q_F$ is the saturation scale in the fundamental representation. The kinematic dependence of $Q_F$ is given phenomenologically by the relation ~\cite{Golec-Biernat:1999qor}
\begin{equation}
	Q_{F}^2(x,\xT)=Q_{p,0}^2\,x^{-\lambda}\,(1-x)^\delta (T_p(\xT)+T_{n}(\xT))\sigma_0\, ,
	\label{eq:q2GBW}
\end{equation}
where $Q_{p,0}^2 = 0.152\,$GeV$^2$, $\lambda=0.215$, $\sigma_0\equiv 2\pi B_G$ denotes the effective transverse area of the nucleon, with $B_G = 0.156$ fm$^2$ . This means that $(T_p(\xT)+T_{n}(\xT))\sigma_0$ counts the density of nucleons per unit transverse area. In this work, we set $\delta=1$ to regulate the large $x$ behavior, noting that the results are almost independent of this choice, as e.g. the results for $\delta=3$ or $5$ would be virtually indistinguishable. 
In what follows, we will drop the fundamental representation index in favor of labeling the nucleus it refers to, e.g. $Q^2_{A/B} $ for nucleus $A/B$. 

In the interest of keeping this derivation simple, we will focus only on the  charge being deposited into the forward region (from projectile $A$ to target $B$), 
corresponding to the first term of \cref{eq:single_quark_density}. The case for the backward deposited charge is, naturally, analogous to these computations and can be obtained by switching nucleus $A$ for nucleus $B$ and inverting rapidity. For the case of forward charge, the quark number density, deposited via multiple scatterings of the collinear quarks of nucleus $A$ with the gluon distributions of nucleus $B$, is then simply described by 
\begin{equation}
	 \left(n_{q_f}\tau\right)_{0,A\to B}=\int\frac{\rmd ^2 \pT}{(2\pi)^2} x_{1}q_{v_f/A}(x_{1},\pT^2,\xT)~D_{\rm fun}(x_2,\xT,\pT)
	 \label{eq:forwardcharge}
	\end{equation}
By using the assumption of isospin symmetry for the proton and neutron PDFs, the $A\to B$ charge is given by
	\begin{equation}
		\begin{split}
			\left(n_u\tau\right)_{0,A\to B}&= \nu_{u,A\to B}\; T_{p,A}(\xT) + \nu_{d,A\to B}\; T_{n,A}(\xT)  \\
			\left(n_d\tau\right)_{0,A\to B}&
			= \nu_{d,A\to B}\; T_{p,A}(\xT) + \nu_{u,A\to B}\; T_{n,A}(\xT)\,, \\
		\end{split}
	\end{equation}
	where we have used \cref{eq:PDFS-pn} to decompose the relation into the individual collinear quark distributions for each quark flavor. We have expressed the resulting integrals as 	
	\begin{equation}
		\nu_{f,A\to B} =\int_0^\infty\frac{\rmd ^2 \pT}{(2\pi)^2}\, x_{1}q_{v_f/p}(x_{1},\pT^2)\,D_{\rm fun}(x_2,\xT,\pT)\,. 
  \label{eq:nuab_nonint}
	\end{equation}
 Now since at any given value of $x$, the dipole distribution is dominated by momenta $\pT^2\sim Q^2_B$, whereas the $\pT^2$ dependence of the collinear quark distribution is only logarithmic due to DGLAP evolution, we can approximate the integral in \cref{eq:nuab_nonint}, by setting $\pT^2=Q_B^2(x_2)$ in the PDF, and approximating the kinematics by 
 \begin{eqnarray}
     x_{1/2} \simeq \frac{Q_{B}(x_2,\xT)}{\sqrt{s_{NN}}} e^{\pm y}
 \end{eqnarray}
 as long as the $x_{1/2}$ dependence is sufficiently mild. By virtue of this approximation, the PDF factorizes and the $\pT$ integration of the dipole gluon distribution yields unity, as the reader can deduce from the definition in \cref{eq:dipole}. We then obtain

\begin{equation}
	\nu_{f,A\to B} \approx  x_{1}q_{v_f/p}(x_{1},Q_B^2)
	\label{eq:NuAB}\,,
\end{equation}
where in the GBW Model, the self-consistent solution to $x_1$ and $x_2$ are given by 
\begin{equation}
	x_1 =\left[\frac{Q_{p,0}^2}{\sNN}T_B(\xT)\,\sigma_0 e^{2y(1+\lambda)}\right]^{1/(2+\lambda)}\,.
	\label{eq:x1forw}
\end{equation}
\begin{equation}
	x_2 =\left[\frac{Q_{p,0}^2}{\sNN}T_B(\xT)\,\sigma_0 e^{-2y}\right]^{1/(2+\lambda)}\,.
		\label{eq:x2forw}
\end{equation}

Next, in order to develop a more analytical feeling to this formulas, we will assume a functional form for the PDFs, following Ref.~\cite{Hou:2019efy}. Since more complex parametrizations primarily affect the large-$x$ regions, we use the simple parametrization
\begin{equation}
	x q_{v_f/p}(x,Q^2 ) = a_0(Q^2)x^{a_{1,f}(Q^2)} \,(1-x)^{a_{2,f}(Q^2)}
\end{equation}
where the exponents $a_{1,f}, a_{2,f}$ and the normalization $a_{0,f}$ typically exhibit a very mild dependence on $Q^2$ (slower than logarithmic) and the exponents $a_{1,f}$ and $a_{2,f}$ are not equal for $u$ and $d$, albeit being close in value. By plugging this functional form into \cref{eq:NuAB}, we get 
\begin{equation}
	\nu_{f,A\to B} \approx  a_0\,x_1^{a_{1,f}}
\end{equation}
which naturally give a parametrical dependence on energy and thickness due to \cref{eq:x1forw}. These dependencies can be explicitly given by 
\begin{equation}
	\nu_{f,A\to B}\sim T_B^{\frac{a_{1,f}}{2+\lambda}} \sNN^{-\frac{a_{1,f}}{(2+\lambda)}}\,\exp\left[2ya_{1,f}\frac{1+\lambda}{2+\lambda} \right]
	\label{eq:parametrical1}
\end{equation}
In high-energy kinematics $\sqrt{\sNN} \gg m_{N}$ the beam rapidity is approximately given by $y_{\rm beam}\approx\log\left(\sNN/m_N^2\right)/2$, such that upon inverting this relationship and plugging it into \cref{eq:parametrical1} we obtain 
\begin{equation}
	\nu_{f,A\to B}\sim T_B^{\frac{a_{1,f}}{2+\lambda}} \,\exp\left[ - \alpha_B (y_{\rm beam} - y(1+\lambda) )\right]
	\label{eq:parametrical2}
\end{equation}
where we define the prefactor inside the exponential as our parametrical \textit{baryon stopping parameter},
\begin{equation}
	\alpha_B\approx\frac{2a_{1,f}}{2+\lambda} \,.
\end{equation}
By inserting the parameters for the GBW model, and the fits to the $x^{a_1}$ tails in the CT18NNLO PDF sets,~\footnote{Since the valence quark distribution is very well constrained, other PDF sets leave the baryon(electric) charge almost unaffected, as it was shown in Ref.~\cite{Garcia-Montero:2023gex}.} we find that $\alpha_B\approx 0.5-0.7$, which is remarkably different from the value of $\alpha_{B/Q}\approx 2$ that one obtains in collinearly factorized pQCD calculations (c.f. \cref{app:collinear}).

We will now validate this approximate expression, by comparing it to a numerical evaluation of the charge deposition in the {\Dipper} framework. 
In \cref{fig:analytical_charge},we present a direct comparison, which has been performed for two different dipole models: on the left we present results for the GBW model, for which the formula is explicitly derived above, while on the right the comparison is presented for the well known IP-Sat model. Indeed, we find that the common features of charge deposition as presented in Ref.~\cite{Garcia-Montero:2023gex} are well reproduced by the analytical formula. The tail  towards midrapidity  and the rapidity shift due to the increasing collisional energies are well described. Nevertheless, for very forward rapidities there is a bigger discrepancy between the phenomenological formula, \cref{eq:NuAB}, and the full result. The  lack of an exponential tail at high rapidities in the parametrical approximation can be traced back to the saddle point approximation leading to \cref{eq:NuAB}. At such high rapidities, transverse momenta at the saturation scale become kinematically prohibited, whereas contributions from modes with $\pT \lesssim Q^2_B$ are still allowed by the integration in the full formula. The rapid decay of the charge deposition in the approximated case is due to the lack of the contribution from these low momentum modes. An additional caveat is that in the far forward region transverse momenta are small due to kinematical constraints. In this case the the perturbative calculation of quark-target scattering is not well controlled and non-perturbative physics associated with hadronization effects can also play an important role. Therefore, the charge deposition in the large rapidity tails ($|\eta|> \eta_{\rm peak}$) should be re-explored with non-perturbative effects in mind.  

By looking at  \cref{eq:parametrical2}, it is easy to see that the overall trend is that denser matter has an antagonistic -and lesser- effect on the charge deposition. This not only means that denser regions of the collision deposit more charges, as expected, but also that denser hotspots move the peak of production closer to midrapidity than more dilute cases. This coincides with the intuitive picture of the quarks being stopped more effectively by a denser wave-packet of glue.

Different solid/dashed lines in  \cref{fig:analytical_charge} correspond to two different isospin configurations for the deposition for the forward charge deposition. Solid lines show the baryon and charge densities for $T_{A}=T_{p,A}= 1 {\rm fm}^{-2}$ (and thus  $T_{n,A}= 0$), while the second configuration of the incoming projectile, shown in dashes, is taken to have only neutron density. These configurations are supposed to roughly mimic the valence quarks of a proton or a neutron scattering through a general target, for which we have chosen  $T_{B}= 1 {\rm fm}^{-2}$, as in \cref{eq:GBWDip} and \cref{eq:q2GBW} the isospin composition of the target is irrelevant. 
We observe that, as expected, the deposition of baryon charge is completely unaffected by the isospin flip of the configuration. On the other had, the electric charge deposited by the quasi $n\rightarrow p$ configuration presents a completely different rapidity profile as the $p\rightarrow p$ configuration due to the different quark content. The picture that emerges is then that in the saturation picture, charges are contained by the valence quarks, and the baryon stopping is performed by the coherent wavepacket in the target. Intuitively, quarks are just being stopped by scattering through a wall of gluons.
This is supported by the fact that diluting of the target $Q_{p,0}^2\rightarrow 0$ shifts the peak to larger rapidities, and reduces the charge deposited at midrapidity. This can be easily seen from~\cref{eq:x1forw}, where decreasing $Q_{p,0}^2$ is effectively the same as increasing $\sNN$.

The right panels of \cref{fig:analytical_charge} explore a comparison of the approximated formula to the IP-Sat model~\cite{Kowalski:2003hm}.
In the IP-Sat model, however, the dipole function is Gaussian only at small transverse momentum, and in general, non-gaussianities affect the deposition of charge and energy. However, while the saturation scale is not explicitly defined in the IP-Sat model as in the GBW model, the gluon distribution is indeed highly peaked at a value $p_\perp^*$, which is the requirement for the computation of the approximated formula. By following standard procedure \cite{Mantysaari:2018nng}, an \textit{effective saturation scale} can then be extracted by defining the saturation scale in analogy to the GBW, such that $Q_S$ is given implicitly by\footnote{The data presented in \cref{fig:analytical_charge}, for the extracted $Q^2_S$, and in the following section corresponds to computations of the full {\Dipper} using the IP-Sat model for the dipole functions, where we have used the parametrization found in Ref.~\cite{Rezaeian:2012ji}.} 
\begin{equation}
		D_{\rm fun}(x,\xT, Q_S^{-1}(x,\xT)) \equiv e^{-1/4}\,.
		\label{eq:effectiveQS}
	\end{equation}
Even with the non-gaussianities present in the IP-Sat model, the qualitative agreement in \cref{fig:analytical_charge}, is still remarkable; in particular the description of the small $\eta_s$ tail is still completely captured by the approximate formula. Because this approximation only depends on the dipole being a highly peaked distribution, we propose that the general parametrization in \cref{eq:param_form_1,eq:param_form_2} can be used for baryon and electric charge depositions in heavy ion collisions, and only needs to be supplemented with a suitable scheme to provide the effective saturation scale $Q_s(x,T(\xT))$ as a function of momentum fraction $x$ and nuclear thickness $T(\xT)$.

\section{Phenomenological implications of the charge deposition}
\label{sec:Results}
Now that we have described the underlying physical picture and obtained a simple analytic description, we continue to perform a more comprehensive analysis of electric and baryon charge deposition at midrapidity in the context of the {\Dipper} model. The analysis here presented will focus on two aspects: collisional energy (beam rapidity)  and system size dependence at midrapidity, which is meant to directly address the recent measurements by STAR and ALICE~\cite{STAR:2008med,STAR:2017sal,ALICE:2013hur,ALICE:2013mez}. In what follows, the determination of the "centrality classes" of the initial state is performed using a proxy for the charged hadron multiplicity at mid-rapidity.Events are separated using the the value of  $\int d^2 \mathbf{x} (\epsilon \tau)^2/3$ at initial time. When matched to hydrodynamics, this quantity corresponds to the initial entropy, which can be used to estimate  the total multiplicity at midrapidity using Eqs. (6) and (7) of Ref.~\cite{Giacalone:2019ldn}.

\subsection{Baryon stopping as a function of center-of-mass energy }
\label{sec:baryon}
\begin{figure*}[t]
	\includegraphics[scale=0.55]{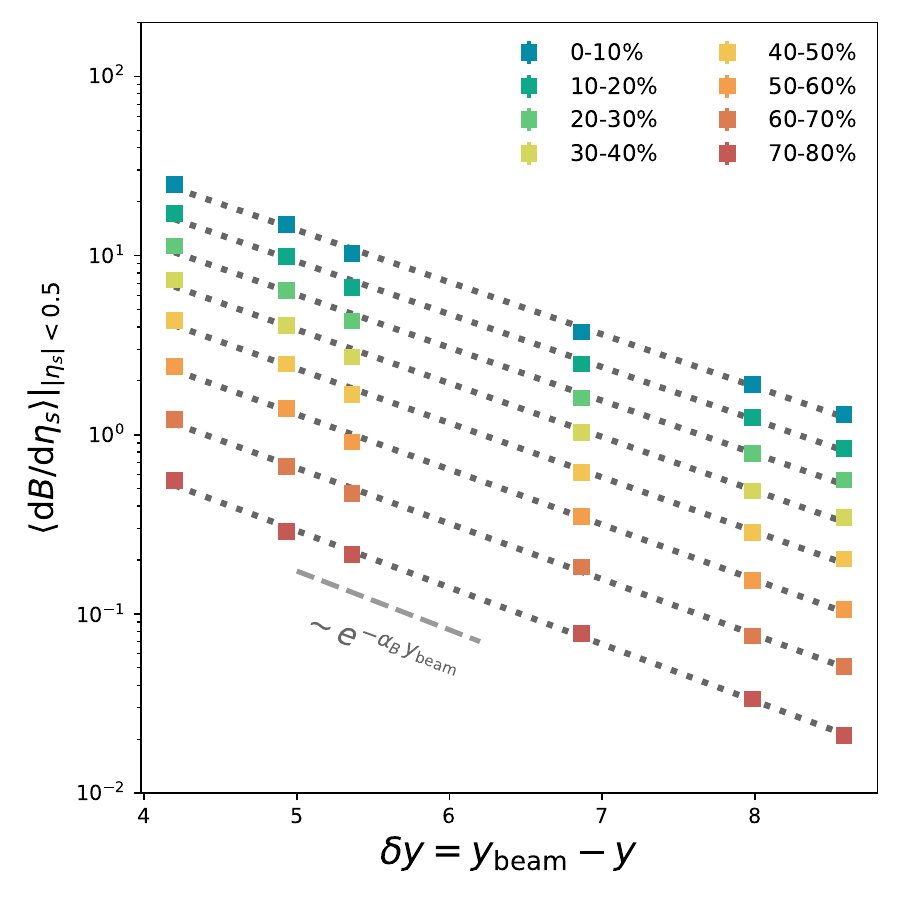}
	\includegraphics[scale=0.55]{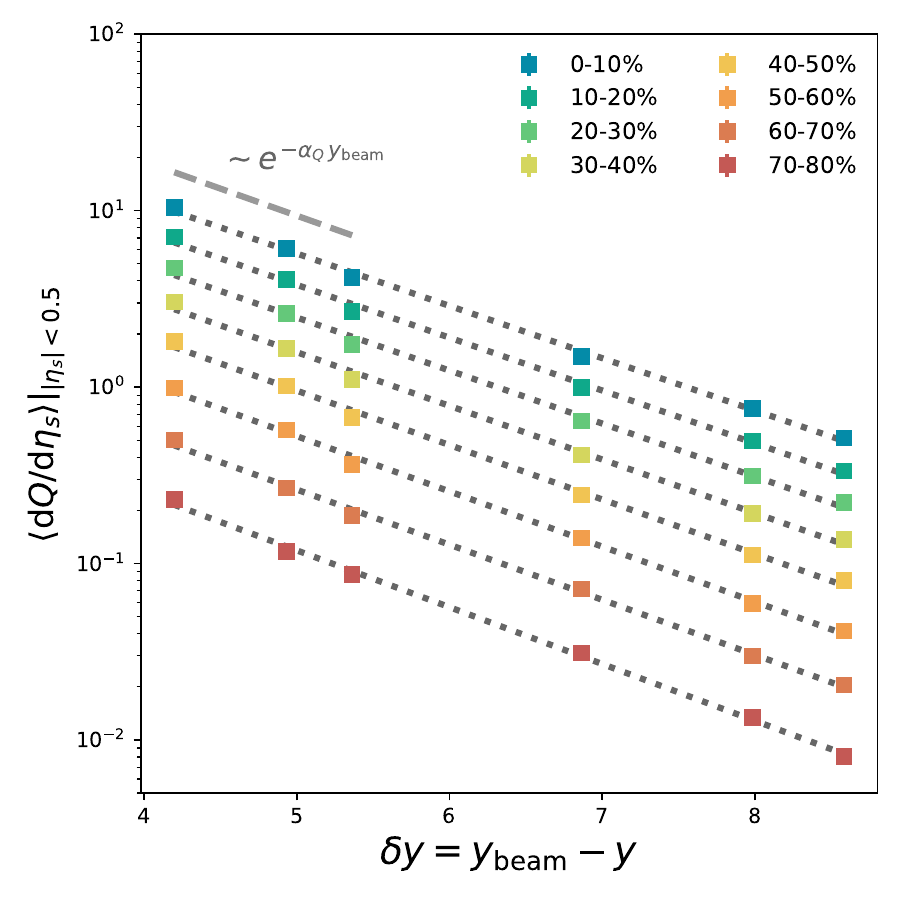}
	\caption{ Midrapidity baryon charge (\textit{left})  and electic charge (\textit{right}) deposition as a function of collisional energy, $\sqrt{s_{\rm NN}}$. The deposited charge follows a power law~$s_{\rm NN}^{\alpha_i/2}$ with $i=[B,Q]$, or equivalent, an exponential with the rapidity shift, $y_{\rm beam}$. }
	\label{fig:charge deposition}
\end{figure*}
In \cref{fig:charge deposition} we present the excitation of function for baryon (\textit{left}) and electric (\textit{right}) charge deposition in Au-Au collisions at midrapidity ($|\eta_s|<0.5$) for a wide energy range, $62.4\,\text{GeV}\leq \SNN \leq 5020\,\text{GeV}$\footnote{For a discussion on the total charge deposited, the reader can refer to Ref.~\cite{Garcia-Montero:2023opu}}. It is important to note that we have fixed the species to $^{197}$Au to avoid including system size effects in baryon deposition which may muddle the extraction of the behavior. 
Additionally, it is worth noting that we have restricted ourselves to energies above $\SNN > 62.4$~GeV to keep the kinematic variables (e.g. $x_2$) for forward deposition inside the range of validity of the CGC EFT. Nevertheless, we note in passing that previous work using the hybrid formalism explores quark deposition at low and intermediate energies (6-60~GeV) \cite{Mehtar-Tani:2011wgy}, and found a good agreement for the position of the peaks of the baryon rapidity density.

We follow previous works~\cite{Arslandok:2023utm,STAR:2024lvy,Lewis:2022arg} and present these results as a function of the rapidity shift, $\delta y = y-y_{\rm beam}$, where we report an exponential decrease of charge deposition with respect to $\delta y$, both for the electric and baryon charge deposition. As it is expected parametrically for midrapidity, 
\begin{equation}
	\left.\frac{\rmd Q_i}{\rmd\eta_s}\right|_{|\eta_s|<0.5} \propto e^{\alpha_i \,(\eta_s - y_{\rm beam})}
	\label{eq:exp_ybeam}
\end{equation}
\begin{figure}[t]
	\includegraphics[scale=0.55]{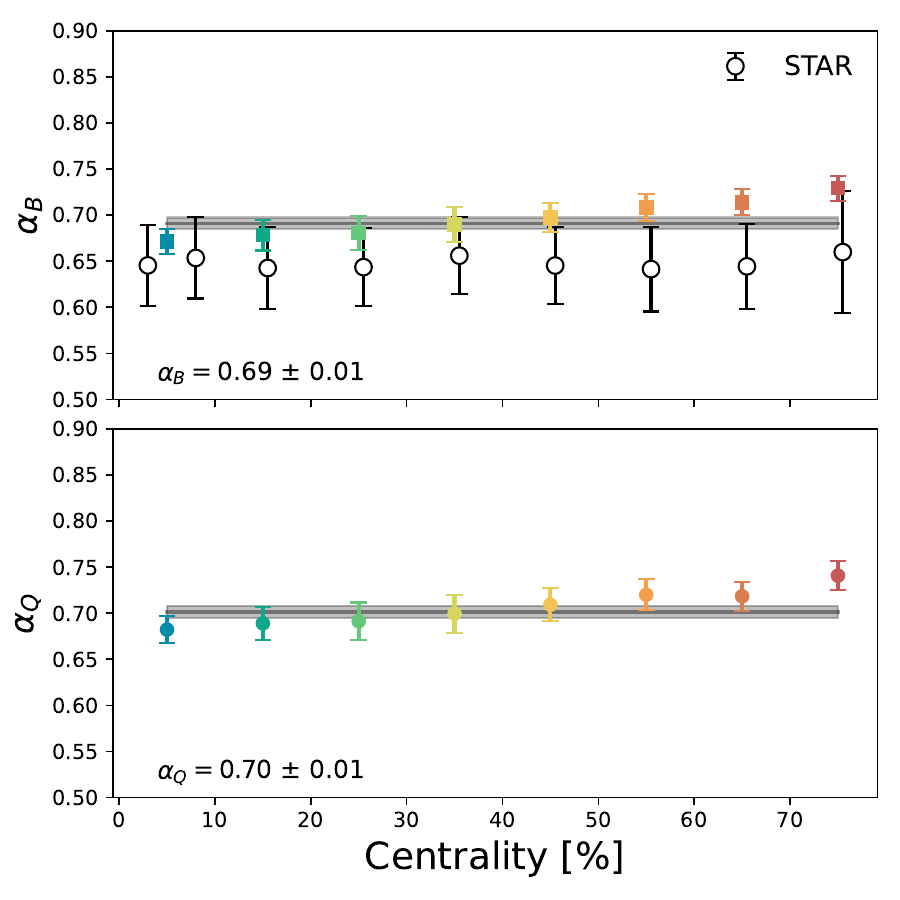}
	\caption{Baryon stopping parameters as extracted from data from the {\Dipper} initial state model as compared to the extracted experimental values from STAR~\cite{STAR:2024lvy}. The gray line corresponds to our theoretical centrality-averaged value. }
	\label{fig:charge deposition_parameters}
\end{figure}
with $Q_i=[B,Q]$. We fit the slope of the computed charges using \cref{eq:exp_ybeam}. The function with the fitted data is given by the gray dotted lines in \cref{fig:charge deposition}.  In \cref{fig:charge deposition_parameters} the reader can find the baryon and electric charge stopping parameters, $\alpha_B$ and $\alpha_Q$ respectively, as a function of the centrality class up to the $70-80\%$. When averaged across all centralitites, they yield 
\begin{equation}
	\alpha_B=0.69\pm 0.01 \quad \text{and} \quad\alpha_Q=0.70\pm 0.01\,,
\end{equation}
which is fully consistent with the semi-analytical result. 
While the numbers presented here are for the initial baryon charge, and not for the final net-proton, as it was presented in Ref.~\cite{Lewis:2022arg}, the baryon stopping parameters computed with the {\Dipper} are quite close to the measured exponents. Because this is such a coarse-grained observable, we also do not expect that the hydrodynamical evolution will exert an important change in the overall trend, albeit this remains to be checked by performing full event-by-event initial state + hydrodynamic +afterburner simulations, as we intend to do in the future.

We finally note that the inclusion of saturation effects in the form of transverse momentum transfer from the target gives rise to an exponent $\alpha_B$, that is quite different from results the values expected for collinear factorisation where $\alpha_B\approx 2$ is much larger (c.f. \cref{app:collinear}). Indeed, it  has been reported that previous implementations of collinear factorization and string dynamics without extra assumptions fails to reproduce the $\sNN$ dependence of the data~\cite{Shen:2022oyg}

\subsection{System size and charge deposition}
\label{sec:systemsize}

\begin{figure}[t]
	\includegraphics[scale=0.55]{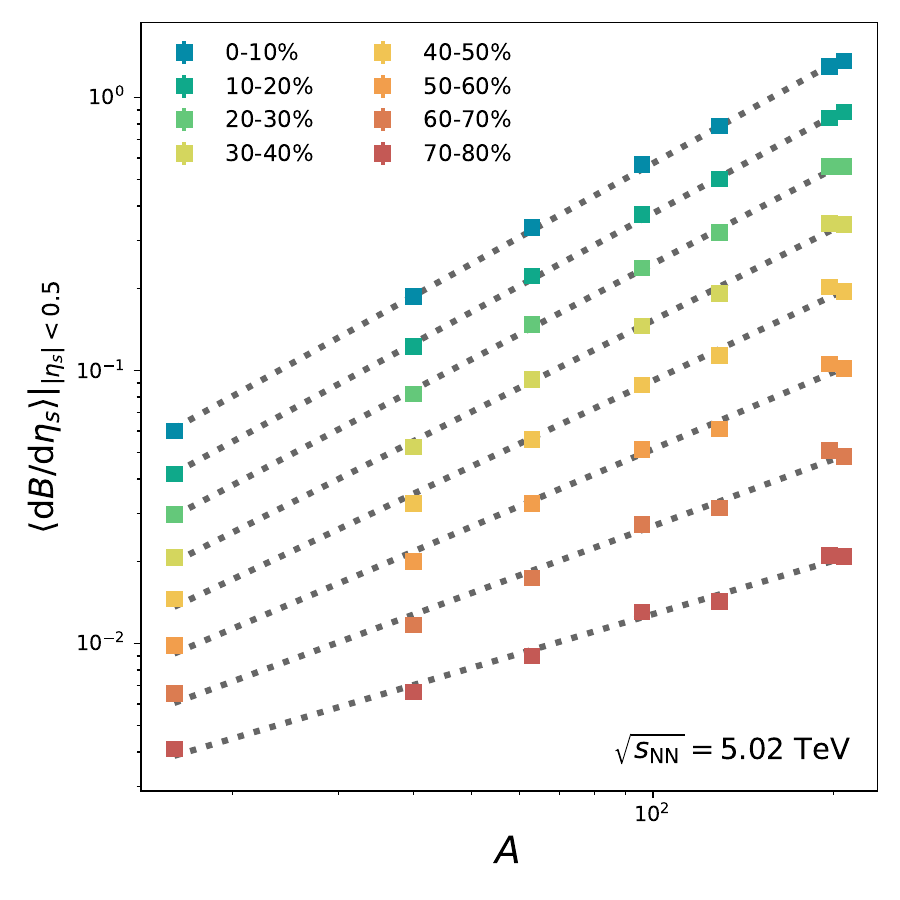}
	\includegraphics[scale=0.55]{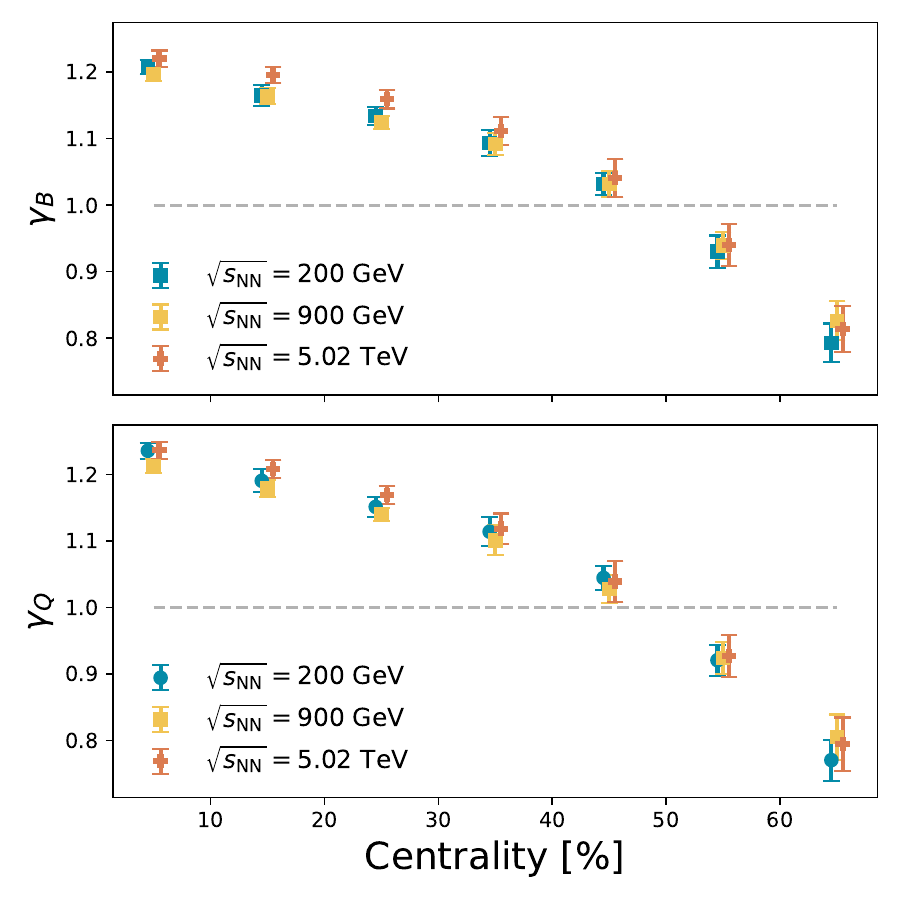}
	\caption{(\textit{top})  Midrapidity baryon charge deposition as a function of nuclear mass number, $A$. The deposited charge follows a power law~$A^{\gamma_i}$ with $i=[B,Q]$.  (\textit{Below})  The baryon deposition parameter $\gamma_i$, found fitted from the {\Dipper}. The parameter $\gamma_i$ is independent of energy, but sensitive to centrality. }
	\label{fig:charge deposition_systems}
\end{figure}
Next, we explore the effect of system size on the deposition of conserved charges. For this, we ran a collection of species, where in order 
to avoid including spurious geometrical effects such as fluctuations due to deformations, we have chosen only species in which a spherical Woods-Saxons profile parametrization is known. The species ran were {\Ox}, {\Ar}, {\Cu}, {\Zr}, {\Xe}, {\Au} and {\Pb}, while varying the collisional energy for $\SNN = [200,900,5020]$~GeV. For simplicity of presentation, we only show the baryon deposition for collisions at $5.02$~TeV in \cref{fig:charge deposition_systems}, but a very similar trend can be found for the other energies. We note that while the baryon charge may be compared across different $A$'s, the electric charge needs to be compared instead to the number of initial protons, $Z$, since the ratio $A/Z$ changes with species. In  \cref{fig:charge deposition_systems}~(\textit{top}) we can observe that the deposited baryon charge follow an increase with increasing system size $A$, which can be represented as 
\begin{equation}
	\left.\frac{\rmd B}{\rmd\eta_s}\right|_{|\eta_s|<0.5} \approx A^{\gamma_{B}}  \qquad \text{and} \qquad 	\left.\frac{\rmd Q}{\rmd\eta_s}\right|_{|\eta_s|<0.5} \approx Z^{\gamma_{Q}} \,.
	\label{eq:system_size_fit_function}
\end{equation}

We have fitted the data to \cref{eq:system_size_fit_function}. The exponents $\gamma_{Q_i}$ can be found in \cref{fig:charge deposition_systems}~(\textit{below}) for different centralities and collisional energies. The exponents are consistent with each other across collisional energies. This is a consequence of the isospin symmetry which we have introduced in \cref{eq:PDFS-pn-iso}.


\section{Conclusions}
\label{sec:conclusions}
We have presented a simplified formula for quark density deposition based on the hybrid formalism of the CFC EFT. The formula samples quarks from collinear PDFs while the scales at which they are sampled are set by the properties of the unintegrated gluon distributions of the nucleus being probed by the traversing valence quark. We found that the formula gives remarkably good results for small and intermediate rapidities, and only deviates from the full result in the fragmentation region very close to the beam rapidity. 

We also presented a phenomenological study on the initial baryon and electric deposition of the {\Dipper}, where we have studied the collisional energy and system size dependence of the deposited charge, for a variety of centralities. We show that the $\SNN$ dependence is consistent with the dependence extracted from experiment~\cite{Lewis:2022arg} at  intermediate energies. 
On the other hand, the study of charge deposition with respect to system size shows that baryon and electric charge deposited in the midrapidity increase as a power law with the total nucleon number ($A^{\gamma_{B/Q}}$).  Nevertheless, the specific power  $\gamma_{B/Q}$ presents an extra centrality dependence.

Since the {\Dipper} provides a rather successful description of experimental results of baryon stopping, which is challenging to describe for other theoretical models, this supports the idea that our physical picture, where baryon stopping in heavy-ion collisions arises due to the stopping of valence quarks by a saturated target, is actually appropriate to describe the process close to mid-rapidity at RHIC and LHC energies. Within this picture, the saturated target provides an effective scale, namely the saturation scale $Q_B$, which exhibits a characteristic center-of-mass energy and nuclear density dependence, that appears to be  consistent with observations from nature. 

 Based on the apparent success of this model it would be interesting to further extend the framework to include additional fluctuations of the charge deposition that arise from event-by-event realizations of the quark content of the colliding hadrons. It would also be interesting, if challenging, to make use the NLO corrections in the CGC EFT, including the $q\to q+g$ NLO channel~\cite{Blaizot:2004wv} and the $g+g\to q\bar{q}$ channel \cite{Dumitru:2005gt} and extend the present theoretical calculations beyond leading order accuracy.
 

\section*{Acknowledgements}
OGM would like to thank Nicole Lewis, Farid Salazar, David Frenklah and Travis Dore for productive discussions. This work is supported by the Deutsche Forschungsgemeinschaft (DFG, German Research
Foundation) through the CRC-TR 211 ‘Strong-interaction matter under extreme conditions’ — project number 315477589 — TRR 211. OGM and SS acknowledge also support by the German Bundesministerium für Bildung und Forschung (BMBF) through Grant No. 05P21PBCAA.

 \appendix 
 \section{Beam-energy dependence of baryon stopping in collinear factorization}
\label{app:collinear}
Below we use a dimensional analysis to obtain the center-of-mass energy $(\sqrt{s}_{NN})$ dependence of the net-charge deposition in collinearly factorized pQCD calculations at high energies. We first note that at leading order, quark charges are deposited via single scattering of collinear partons, and we consider a generic $h_1 h_2 \to q_f X$ process, for which the cross-section is given by a convolution of the relevant PDFs $f_{i/h_j}(x,\mu)$ and the cross-section for the partonic sub-process as 
\begin{equation}
\begin{split}
	\frac{\rmd \sigma^{h_1 h_2\to q_f X}}{\rmd^2 p_T \rmd y_q } =\sum_{ij}\int \rmd x_1\, \rmd x_2& \, f_{i/h_1}(x_1,\mu) \, f_{j/h_2}(x_2,\mu) \\
& \times \frac{\rmd \sigma^{ij\to q_f X}}{\rmd x_1 \rmd x_2\rmd^2 p_T \rmd y_q }
 \label{eq:quarknaivecs}
 \end{split}
\end{equation}
where $p_T^{q}$ and $y^{q}$ denote the transverse momentum and rapidity of the outgoing quark. Based on the cross-section for this process, we can then extract the net-quark yield by normalizing with the total inelastic cross section, 
\begin{equation}
\begin{split}
	\frac{\rmd N^{h_1 h_2\to q_f X}}{\rmd y_q } =\frac{1}{\sigma_{\rm inel}(\sNN)}\int \rmd^2 p_T \frac{\rmd \sigma^{h_1 h_2\to q_fg}}{\rmd^2 p_T^{q} \rmd y_q } 
 \end{split}
\end{equation}
By kinematics the momentum fractions $x_1$ and $x_2$ determine the ratio of the hadronic center of mass energy $\sqrt{s_{NN}}$ to the center of mass energy  of the partonic sub-process $\sqrt{s}$ as
\begin{equation}
s=x_1\,x_2\,\sNN 
\label{eq:wurstsalat}
\end{equation}
Now the differential cross-section for the partonic sub-process $\frac{\rmd \sigma^{ij\to q_f X}}{\rmd x_1 \rmd x_2\rmd^2 p_T^{q} \rmd y_q }$, is determined by the dimensionless variables $x_1$,$x_2$,$y_q$ and the energy scales $\sqrt{s}$ and $p_T^{q}$. In the limit where both of these scales $p_T^{q},\sqrt{s} \gg \Lambda_{\rm QCD}$, dimensional analysis suggests that, up to an overall pre-factor $1/s^2$ the cross-section can be expressed as in terms of a dimensionless function $\frac{\rmd \tilde{\sigma}^{ij\to q_f X}}{\rmd x_1 \rmd x_2\rmd^2 p_T \rmd y_q }\left(\frac{p_T^{q}}{\sqrt{s}}\right)$ of the ratio of these two scales, i.e.
\begin{equation}
\frac{\rmd \sigma^{ij\to q_f X}}{\rmd x_1 \rmd x_2\rmd^2 p_T^{q} \rmd y_q } =\frac{1}{s^2}\frac{\rmd \tilde{\sigma}^{ij\to q_f X}}{\rmd x_1 \rmd x_2\rmd^2 p_T \rmd y_q }\left(\frac{p_T^{q}}{\sqrt{s}}\right)
\end{equation}
By inserting this expression into Eq.~(\ref{eq:quarknaivecs}), we can absorb one power of $1/s$ to change the transverse momentum $p_T^{q}$ integration, to an integration over the dimensionless variable $\frac{p_T^{q}}{\sqrt{s}}$, and use Eq.~(\ref{eq:wurstsalat}) to express 

\begin{widetext}
\begin{equation}
	\frac{\rmd N^{h_1 h_2\to q_f X}}{\rmd y_q } =\frac{1}{\sigma_{\rm inel}(\sNN)\sNN}\int 
\frac{\rmd x_1}{x_1}f_{i/h_1}(x_1,\mu) 
\int \frac{\rmd x_2}{x_2}  \, f_{j/h_2}(x_2,\mu) \int\rmd^2  \left(\frac{ p^q_T}{\sqrt{s}}\right) \frac{\rmd \tilde{\sigma}^{ij\to q_f X}}{\rmd x_1 \rmd x_2\rmd^2 p_T \rmd y_q }\left(\frac{p_T^{q}}{\sqrt{s}}\right)
\end{equation}
\end{widetext}
In the high energy limit, $\sNN \approx m_N^2 \exp[2\,y_{\rm beam}]$. 
Additionally, it is parametrically known that $\sigma_{\rm inel}(\sNN)\approx \log^2(\sNN)\sim y_{\rm beam}^2 $. Baryon and electric charge can be then obtained by computing the some process for the antiquark and take the appropriate linear combination. The resulting exponent is then 
\begin{equation}
\alpha_B\sim 2
\end{equation}
which is qualitatively compatible to the event generator results discussed in Ref.~\cite{Lewis:2022arg}.

A small caveat is in order: this result is valid as a very coarse approximation, but corrections will not change the qualitative behaviour $\alpha_B>1$. The difference from this result to the once presented in the main text rest on the difference of kinematics. While here the scaling is solely driven by the momentum conservation of one scattering, the saturation results are prompted by the total momentum broadening of a charge through a medium which is described by a distribution with characteristic scale $Q_S$. Therefore, the so-called naive quark stopping comprises a very particular limit of the quark stopping picture.

\bibliography{References}
\end{document}